# Robust Sensor Fusion for Indoor Wireless Localization

**Gang Wang***
Center for Robotics, School of Information and Communication Engineering
University of Electronic Science and Technology of China
No.2006, Xiyuan Ave, West Hi-Tech Zone, 611731, Chengdu, Sichuan, P.R.China
wanggang_hld@ uestc.edu.cn

**Zuxuan Zhang**
School of Information and Communication Engineering
University of Electronic Science and Technology of China
No.2006, Xiyuan Ave, West Hi-Tech Zone, 611731, Chengdu, Sichuan, P.R.China
changtsuhsuan @qq.com

**ABSTRACT**

Location knowledge in indoor environment using Indoor Positioning Systems (IPS) has become very useful and popular in recent years. Indoor wireless localization suffers from severe multi-path fading and non-line-of-sight conditions. This paper presents a novel indoor localization framework based on sensor fusion of Zigbee Wireless Sensor Networks (WSN) using Received Signal Strength (RSS). The unknown position is equipped with two or more mobile nodes. The range between two mobile nodes is fixed as priori. The attitude (roll, pitch, and yaw) of the mobile node are measured by inertial sensors (ISs). Then the angle and the range between any two nodes can be obtained, and thus the path between the two nodes can be modeled as a curve. Through an efficient cooperation between two or more mobile nodes, this framework effectively exploits the RSS techniques. This constraint help improve the positioning accuracy. Theoretical analysis on localization distortion and Monte Carlo simulations shows that the proposed cooperative strategy of multiple nodes with extended Kalman filter (EKF) achieves significantly higher positioning accuracy than the existing systems, especially in heavily obstructed scenarios.

**KEYWORDS:** Indoor Localization, Wireless Localization, Node Cooporation, Sensor Fusion, Received Signal Strength (RSS), Extended Kalman Filter (EKF)

## 1  INTRODUCTION

The success of outdoor positioning and applications based on the global positioning systems (GPS) provides an incentive to the research and development of indoor positioning system. Although the GPS provides good solutions to outdoor positioning, it has limited performance for indoor environments. Infrared, Radio Frequency (RF), and Ultra Sound signals are major technologies for indoor positioning systems (Angelis et al. ,2013),(Colombo et al. ,2014),(Zhang et al. ,2013),(Hightower et al. ,2014). Unlike outdoor areas, indoor geolocation faces severe technical challenges (Pahlavan et al. ,2002),(Chan and Baciu,2012),(Quyang et al. ,2010). The indoor environment imposes different challenges on location discovery due to the dense multipath effect and building material dependent propagation effect. Thus, an in-depth understanding of indoor radio propagation for positioning is crucial for efficient design and deployment.

The existence of radio connectivity and the attenuation of radio signal with distance are attractive properties that could potentially be exploited to estimate the positions of small wireless devices featuring low-power radios. Received signal strength (RSS), a standard feature in most radios, has attracted a lot of attention in the recent literatures. Given a model of radio signal propagation in a building or other environment, RSS can be used to estimate the distance from a transmitter to a receiver, and thereby estimate the position of a mobile node (MN). However, this approach requires detailed models of RF propagation and does not account for variations in receiver sensitivity and orientation. RSS eliminates the need for additional hardware in small wireless devices, and exhibits favorable properties with respect to power consumption, size and cost.

In order to tackle the challenges, with the advances of modern technologies, several studies have shown the benefits from sensor fusion between heterogeneous set of machines. Few papers are focused on sensor fusion of multiple mobile nodes in a cooperation way to tackle the positioning accuracy problem. This motives us to put forward new idea for localization. This paper presents a novel localization framework for multiple mobile nodes based on sensor fusion of Zigbee and Inertial sensors (IS). The unknown position is equipped with two or more mobile nodes. The range between two mobile nodes is fixed as priori. The attitude (roll, pitch, and yaw) of the mobile node are measured by inertial sensors (ISs). Then the angle and the range between any two nodes can be obtained. . With the help of two or more mobile nodes, the performance will be improved.

In this paper, we inquired about the RSS solutions on indoor localization, and proposed a new RSS-based algorithm. The rest of this paper is organized as follows. In Section 2, the novel localization framework is introduced. In Section 3, theoretical analysis is shown how the approach improved the positioning performance. The Kalman filtering solution to the proposed approach is shown in Section 4. The simulations are described and analyzed in Section 5. Conclusions are given in Section 6.

## 2    THE NOVEL LOCALIZATION FRAMEWORK

Radio channel characterization in a specific environment is obtained by looking for a mathematical relationship between the RSS value, usually in dBm, and the distance between the mobile and the reference node.

The IEEE802.15.4 protocol has given the simplified wireless channel transmission model:

$$RSSI(d) = \begin{cases} Pt - 40.2 - 10*2*\lg(d), d \leq 8m \\ Pt - 58.5 - 10*3.3*\lg(d), d > 8m \end{cases} \quad (1)$$

where $Pt$ is the signal transmission power, usually is -0.1dBm. $d$ is the transmitter_receiver distance between two nodes. It can be seen that the attenuation of signal strength with distance is the index relationship.

The average Path Loss of the radio signal propagation model that means the relationship between received power and distance is formed as:

$$RSS(d) = RSS(d_0) + 10n \log \frac{d}{d_0} + X_{dBm} \quad (2)$$

where $n \in [2,4]$ is the attenuation factor (or loss exponent) of the specific environment and $n = 2$ for free space. The $RSS(d_0)$ is the received power from the transmitter at a known close distance $d_0$ and $X_{dBm}$ denotes a zero mean Gaussian random variable that reflect the interference such as Shadowing, multi-path and diffusion phenomenons from indoor environment.

In reality, the real received signal power would not conform to the Path Loss model. With Eq. (3) we can obtain a relation between transmitter and receive power, simplified for the case of a 1-meter reference distance as:

$$RSS = A + 10n \log d + X_{dBm} \quad (3)$$

We computed the $A$ and $n$ values by collecting RSS values between two nodes at different distances, and performed a logarithmic interpolation of RSS data according to Eq. (3). The computed values of A and n were $A = -37.3420$ $n = 1.9236$ and the standard deviation of noise is $\sigma = 3.0130$, results are quite consistent with the fact that we considered an indoor scenario. As the same, the IEEE802.15.4 model's values is $A = -40.3$ $n = 2$ and $\sigma = 2.3662$

To help illustrate the idea, here we consider reference nodes (RN) placed in regular grids, such as four reference nodes at the corners of a square grid with known coordinates. The proposed strategy, however, generalizes to non-regular grids of any shape, and to either three, four, five or more references nodes working together. Each sensor is equipped with simple hardware, such as a Zigbee radio, a single omni-directional antenna and an inexpensive clock.

The range between two mobile nodes are fixed as priori. The attitude (roll, pitch, and yaw) of the mobile node are measured by inertial sensors (ISs). Then the angle and the range between two nodes can be obtained.

In the first phase, mobile node 1 (MN1) broadcasts a beacon packet. The four reference nodes record the respective Signal Strength (SS) of the beacon packet, and at the same time, mobile node 2 (MN2) performs the same process as MN1.

In the second phase, the four reference nodes send the recorded SS information to the MN1 and MN2, respectively.

Finally, both of the MNs pass all RSS measurements to a central node (CN), the central node send the information to the PC. Then, the PC takes advantage of the positional relation and the RSS to perform a joint estimation of both sensors' locations.

Due to the wireless broadcast advantage, MN1 obtains the RSS of its buddy, MN2, for free. Hence, MN2 does nothing more or different than the conventional non-cooperative case. No additional communication or computation overhead is required at the sensors side.

Suppose there is a signal receiver (SR) in each mobile node. Denote the ith signal receiver position in the ith mobile node by $\mathbf{x}_i = (x_i, y_i, z_i), i = 1, 2, \cdots, M$, where $M$ is the number of mobile nodes. Denote the jth reference node position by $\mathbf{X}_j = (X_j, Y_j, Z_j), j = 1, 2, \cdots, N$, where $N$ is the number of available reference nodes. Then the RSS, $RSS_{ij}$, from the ith mobile node to jth reference node is given as:

$$RSS_{ij} = A + 10n\log_{10}(\|\mathbf{x}_i - \mathbf{X}_j\|_2) + e_{ij}$$
$$\mathbf{x}_i = (x_i, y_i, z_i) \qquad \mathbf{X}_j = (X_j, Y_j, Z_j) \qquad (4)$$
$$\|\mathbf{x}_i - \mathbf{X}_j\|_2 = \sqrt{(x_i - X_j)^2 + (y_i - Y_j)^2 + (z_i - Z_j)^2}$$

and $e_{ij}$ represents disturbances or model errors caused by the system, mainly including the multi-path effects.

It should be mentioned in model as Eq.(4) that only the position of the mobile nodes is to estimate. In Eq. (4), $RSS_{ij}$ and $(X_j, Y_j, Z_j)$ can be obtained by signal receivers, and then there are three variables, $x_i$, $y_i$, $z_i$, left unknown. Thus at least three reference nodes are necessary for the above solution. The solution is influenced by model errors $e_{ij}$, which has an order of 5m. In the presence of differential RSS (DRSS), the range, $RSS_{0j}$, from the signal receiver with known position $\mathbf{x}_0 = (x_0, y_0, z_0)$ to jth reference node is shown as:

$$RSS_{0j} = A + 10n\log_{10}(\|\mathbf{x}_0 - \mathbf{X}_j\|_2) + e_{0j} \qquad (5)$$

Since $e_{ij}$ and $e_{0j}$ have the same kind of model errors if the ith signal receiver is near to the known position signal receiver, differencing Eqs. (4) -(5) gives:

$$RSS_{ij} - RSS_{0j} = 10n\log\|\mathbf{x}_i - \mathbf{X}_j\|_2 - 10n\log\|\mathbf{x}_0 - \mathbf{X}_j\|_2 + (e_{ij} - e_{0j}) \qquad (6)$$

where the left errors $e_{ij} - e_{0j}$ only has an order of 1m in practice, and can be approximately modeled as Gaussian white noise with zeros-means and a typical standard deviation of 1m (Andersen et al. ,1995). The obtained accuracy by DRSS is increased by differencing the model errors.

The fusion of IS and RSS, denoted by IS-RSS, motives us put forward a new sensor fusion framework to deal with the above two problems. The new framework is shown in Figure 1, where IS-RSS is further fused. The unknown position is equipped with two or more mobile nodes. The range between any two mobile nodes is fixed as priori. The attitude (roll, pitch, and yaw) of the mobile node are measured by inertial sensors (ISs). Then the angle and the range between any two nodes can be obtained. Thus the path between the two nodes can be modeled as a curve.

It is can be expected that the available number of reference nodes can be reduced with the increase of the number of mobile nodes.

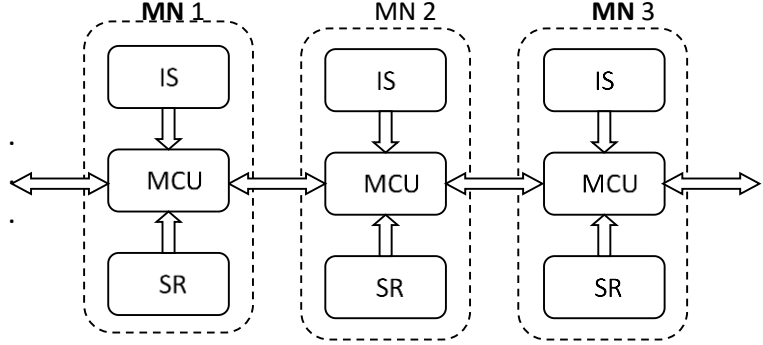

Figure 1: Framework of proposed scheme.

## 3 THEORETICAL ANALYSIS

In this section, the three-dimension (3D) conditions are considered.

For simplicity, suppose the 1st signal receiver as host mobile node, which receive the RSS information from the jth ( $j = 1, 2, \cdots, N$ ) reference node. So does other mobile nodes. Then the curve between the 1st signal receiver and the ith signal receiver can be expressed by an intersection of two surfaces as follows:

$$\begin{cases} S_1(x,y) = 0 \\ S_2(x,y) = 0 \end{cases} \tag{7}$$

The most popular curve between two MNs is straight lines. The ith position $x_i = (x_i, y_i, z_i), i > 1$ can be expressed with the 1st position $x_1 = (x_1, y_1, z_1)$ as follows:

$$\begin{cases} x_i = f_{i1}(x_1, y_1, z_1) \\ y_i = f_{i2}(x_1, y_1, z_1) \end{cases}, \quad i > 1. \tag{8}$$

Denote by

$$\begin{aligned} \mathbf{f}_i &= [f_{i1}, f_{i2}]^T \\ f_{i1} &= f_{i1}(x_1, y_1, z_1), \quad i > 1 \\ f_{i2} &= f_{i2}(x_1, y_1, z_1) \end{aligned} \tag{9}$$

Then the $RSS_{1j}$ from the 1st signal receiver to jth reference nodes is given as:

$$RSS_{1j} = A + 10n\log_{10}(\|\mathbf{x}_1 - \mathbf{X}_j\|_2) + e_{1j} \tag{10}$$

Substituting Eq.(9) into Eq.(3) yields

$$RSS_{ij} = A + 10n\log_{10}(\|\mathbf{f}_i - \mathbf{X}_j\|_2) + e_{ij} \tag{11}$$

With the increase of the number of the reference nodes, more equations like (10) and (11) are obtained.

Then the curve between the 1st signal receiver and the ith signal receiver can be expressed by an intersection of two surfaces as follows:

$$\begin{cases} S_1(x,y,z) = 0 \\ S_2(x,y,z) = 0 \end{cases} \quad (12)$$

The most popular curve between two MNs is straight lines. With the help of obtained range and angle by IS, the ith position $x_i = (x_i, y_i, z_i), i > 1$ can be expressed with the 1st position $x_1 = (x_1, y_1, z_1)$ as follows:

$$\begin{cases} x_i = f_{i1}(x_1, y_1, z_1) \\ y_i = f_{i2}(x_1, y_1, z_1), \quad i > 1. \\ z_i = f_{i3}(x_1, y_1, z_1) \end{cases} \quad (13)$$

Denote by

$$\begin{aligned} \boldsymbol{f}_i &= [f_{i1}, f_{i2}, f_{i3}]^T \\ f_{i1} &= f_{i1}(x_1, y_1, z_1) \\ f_{i2} &= f_{i2}(x_1, y_1, z_1), \quad i > 1 \\ f_{i3} &= f_{i3}(x_1, y_1, z_1) \end{aligned} \quad (14)$$

Then the $RSS_{1j}$ and $RSS_{ij}$ is the same as Eqs. (9) - (10).

Without losing generality, a curve including position $x_i = (x_i, y_i, z_i), i = 1, 2, \cdots, M$ can be modeled as

$$\begin{cases} y_i = S_{1i}(x_i) \\ z_i = S_{2i}(x_i) \end{cases} \quad (15)$$

## 4 FILTERING SOLUTION

In order to make a tradeoff between performance and computation complexity, EKF is utilized here to estimate the localization information for comparisons.

In practice, the curve model as Eqs.(9)(14) between two nodes are obtained after getting their positions, so it cannot be directly utilized by direct solution and it will be better estimated in EKF technique by state augmentations.

### 4.1 EKF Solution for 2D tracking

We first introduce EKF for 2D tracking when the curve model between two nodes is not considered. As mentioned above, suppose the 1st signal receiver as MN1, which receive the RSS information from the jth ($j = 1, 2, \cdots, N$) reference node. The ith position $(x_i, y_i, z_i)$ is expressed with the 1st position. Then for the case of $N$ available reference nodes, the set of equations for 3D RSS are (10) and (11). Denote by

$$d_{ij} = A + 10n\log_{10}(\|\boldsymbol{f}_i - \boldsymbol{X}_j\|_2)$$

where

$$\boldsymbol{f}_1 = \begin{cases} f_{11} = x_1 \\ f_{12} = y_1 \\ f_{13} = z_1 \end{cases} \text{ and } \boldsymbol{f}_i = \begin{cases} f_{i1} = x_i \\ f_{i2} = y_i \\ f_{i3} = z_i \end{cases}, \quad i > 1$$

Then (9) and (10) can be further combined as:

$$RSS_{ij} = d_{ij} + e_{ij} \tag{16}$$

where $i = 1, 2, \cdots, M$; $j = 1, 2, \cdots, N$.

Denote by:

$$\begin{aligned} \boldsymbol{RSS} &= [RSS_{11}, \cdots, RSS_{1N}, RSS_{21}, \cdots, RSS_{2N}, \cdots, RSS_{M1}, \cdots, RSS_{MN}]^T_{MN \times 1} \\ \boldsymbol{D} &= [d_{11}, \cdots, d_{1N}, d_{21}, \cdots, d_{2N}, \cdots, d_{M1}, \cdots, d_{MN}]^T_{MN \times 1} \\ \boldsymbol{E} &= [e_{11}, \cdots, e_{1N}, e_{21}, \cdots, e_{2N}, \cdots, e_{M1}, \cdots, e_{MN}]^T_{MN \times 1} \end{aligned} \tag{17}$$

The RSS from the jth reference node can be written as:

$$\boldsymbol{RSS} = \boldsymbol{D} + \boldsymbol{E} \tag{18}$$

Recursive Bayesian estimation is a general probabilistic approach for sequentially estimating an unknown state probability density function over time using incoming noisy measurements and a mathematical process model. The problem can often be cast in terms of estimating the state of a discrete-time nonlinear dynamic system:

$$\begin{cases} \boldsymbol{x}_{k+1} = f(\boldsymbol{x}_k, u_k) \\ \boldsymbol{z}_k = h(\boldsymbol{x}_k, v_k) \end{cases} \tag{19}$$

where $\boldsymbol{x}_k$ represents the unobserved state of the system and is the sensor observations. The process noise $u_k$ drives the dynamic system, and the observation noise is given by $v_k$. Note that we are not assuming additivity of the noise sources. The system dynamic model $f(\cdot)$ and observation model $h(\cdot)$ are assumed known.

We define the state vector :

$$\boldsymbol{x} = [x, y, \dot{x}, \dot{y}]^T \tag{20}$$

Corresponding to 2D position and velocity for tracking purposes. The velocity is modeled as a random-walk process under near constant, then:

$$\begin{bmatrix} \dot{x}(k) \\ \dot{y}(k) \\ \ddot{x}(k) \\ \ddot{y}(k) \end{bmatrix} = \begin{bmatrix} 0 & 0 & 1 & 0 \\ 0 & 0 & 0 & 1 \\ 0 & 0 & 0 & 0 \\ 0 & 0 & 0 & 0 \end{bmatrix} \begin{bmatrix} x(k) \\ y(k) \\ \dot{x}(k) \\ \dot{y}(k) \end{bmatrix} + \begin{bmatrix} 0 \\ 0 \\ w_{vx} \\ w_{vy} \end{bmatrix} \tag{21}$$

where $w_{vx}$ and $w_{vy}$ are the corresponding process noises. Then the state-space equation of RSS receiver is shown as

$$\dot{\boldsymbol{x}} = \boldsymbol{A}\boldsymbol{x} + \boldsymbol{U} \tag{22}$$

where

$$\boldsymbol{A} = \begin{bmatrix} 0_{2 \times 2} & I_{2 \times 2} \\ 0_{2 \times 2} & 0_{2 \times 2} \end{bmatrix}.$$

$$\boldsymbol{V} = [0, 0, w_{vx}, w_{vy}]^T$$

The $w_{vx}, w_{vy}$ are selected process terms to model the velocity (33). The measurement equation for 2D RSS is given as

$$Z = RSS = h(x) + E \tag{23}$$

Eqs. (21)-(23) shows a continuous-time model of a system, and their discrete-time model is given as:

$$\begin{aligned} x(k+1) &= Fx(k) + W(k) \\ R(k+1) &= h(x(k)) + E(k) \end{aligned} \tag{24}$$

where $F = e^{AT_s}$, $W(k) \approx T_s V$ and $T_s$ is the sampling period. $W(k)$ and $E(k)$ are zero-mean norm distributed white noise and are characterized by covariance matrices $Q_W$ and $Q_E$, respectively. Then the corresponding EKF on-line solutions are as follows:

$$\begin{aligned} \hat{x}^-(k) &= F\hat{x}^+(k-1) \\ P^-(k) &= FP^+(k-1)F^T + Q_w \\ K(k) &= P^-(k)H^T(k)\left[H(k)P^-(k)H^T(k) + Q_E(k)\right]^{-1} \\ P^+(k) &= \left[I - K(k)H(k)\right]P^-(k) \\ \hat{x}^+(k) &= \hat{x}^-(k) + K(k)\left[R(k) - h(x^-(k))\right] \end{aligned} \tag{25}$$

where

$$H(k) = \frac{\partial h(x)}{\partial x}\bigg|_{x=x^-(k)} = \left[\left[H_{kj}\right]_{MN\times 2} \quad 0_{MN\times 2}\right]\bigg|_{x=x^-(k)} \tag{26}$$

Let $k = (i-1)M + j$, where $i = 1, 2, \cdots, M$, $j = 1, 2, \cdots, N$ and $k = 1, 2, \cdots, MN$. The elements in Eq.(26) can be described as:

$$H_{k1} = \frac{\partial RSS_{ij}}{\partial x(k)} = \frac{10n\log_{10} e \cdot (f_{i1} - X_j)}{d_{ij}^2}, \quad H_{k2} = \frac{\partial RSS_{ij}}{\partial y(k)} = \frac{10n\log_{10} e \cdot (f_{i2} - Y_j)}{d_{ij}^2}$$

### 4.2 EKF Solution for 3D tracking

The position relations between two nodes is:

$$f_1 = \begin{cases} f_{11} = x_1 \\ f_{12} = y_1 \\ f_{13} = z_1 \end{cases} \text{ and } f_i = \begin{cases} f_{i1} = x_i \\ f_{i2} = y_i, \quad i > 1 \\ f_{i3} = z_i \end{cases} \tag{27}$$

The velocity is modeled as Eq. (28), the state vector is expressed as Eq.(29), the state-space equation of RSS receiver is shown as Eq.(30).

$$x = \begin{bmatrix} x(k) & y(k) & z(k) & \dot{x}(k) & \dot{y}(k) & \dot{z}(k) \end{bmatrix}^T \tag{28}$$

$$\begin{bmatrix} \dot{x}(k) \\ \dot{y}(k) \\ \dot{z}(k) \\ \ddot{x}(k) \\ \ddot{y}(k) \\ \ddot{z}(k) \end{bmatrix} = \begin{bmatrix} 0 & 0 & 0 & 1 & 0 & 0 \\ 0 & 0 & 0 & 0 & 1 & 0 \\ 0 & 0 & 0 & 0 & 0 & 1 \\ 0 & 0 & 0 & 0 & 0 & 0 \\ 0 & 0 & 0 & 0 & 0 & 0 \\ 0 & 0 & 0 & 0 & 0 & 0 \end{bmatrix} \begin{bmatrix} x(k) \\ y(k) \\ z(k) \\ \dot{x}(k) \\ \dot{y}(k) \\ \dot{z}(k) \end{bmatrix} + \begin{bmatrix} 0 \\ 0 \\ 0 \\ w_{vx} \\ w_{vy} \\ w_{vz} \end{bmatrix} \quad (29)$$

Where $w_{vx}$, $w_{vy}$ and $w_{vz}$ are the corresponding process noises. Then the state-space equation of RSS receiver is shown as

$$\dot{x} = Ax + U \quad (30)$$

Their discrete-time model is given as:

$$\begin{aligned} x(k+1) &= Fx(k) + W(k) \\ R(k+1) &= h(x(k)) + E(k) \end{aligned} \quad (31)$$

where $i = 2, 3, \cdots, M$, $j = 1, 2, \cdots, N$.

Then the corresponding EKF on-line solutions are as Eq.(25):
where

$$\begin{aligned} H(k) &= \left. \frac{\partial h(x)}{\partial x} \right|_{x=x^-(k)} \\ &= \left. \left[ [H_{kj}]_{MN \times 3} \quad 0_{MN \times 3} \right] \right|_{x=x^-(k)} \end{aligned} \quad (32)$$

Let $k = (i-1)M + j$, where $i = 1, 2, \cdots, M$, $j = 1, 2, \cdots, N$ and $k = 1, 2, \cdots, MN$. The elements in (32) can be described as:

$$H_{k1} = \frac{\partial RSS_{ij}}{\partial x(k)} = \frac{10n \log_{10} e \bullet (f_{i1} - X_j)}{d_{ij}^2}, \quad H_{k2} = \frac{\partial RSS_{ij}}{\partial y(k)} = \frac{10n \log_{10} e \bullet (f_{i2} - Y_j)}{d_{ij}^2}, \quad H_{k3} = \frac{\partial RSS_{ij}}{\partial z(k)} = \frac{10n \log_{10} e \bullet (f_{i3} - Z_j)}{d_{ij}^2}.$$

## 5　SIMULATIONS AND EXPERIMENTS

In the experiments, EKF algorithm is utilized to test the feasibility of the proposed method. Suppose there are two nodes, which equipped with signal receivers and inertial sensors, form a formation.

The reference nodes are attached at fixed positions and communicate with mobile nodes. The mobile node was moving around freely. But in test, a number of trials were conducted followed a predefined path. The mobile nodes receive the message signal from the reference nodes. Then, the mobile nodes will extract the signal strength. After that, the signals are sent to location server. The location server calculates the location of mobile node and shows the results.

The network comprises of some ZigBee nodes in which there are reference nodes and mobile nodes. The CC2431 chips stand for the mobile nodes and the CC2430 chips stand for the reference nodes.

One of the reference nodes is connected to a laptop computer via a RS232-to-USB serial interface and the node is the ZigBee coordinator (ZC). The rest of the nodes in the network are implemented as ZigBee routers (ZR). The implemented localization system can be categorized as a centralized architecture as described in ZigBee Telecom Application Profile Specification. The two-dimensional locations in this system are identified using Cartesian coordinate system or pair of numerical coordinates.

An actual position is represented by the coordinate $(x, y)$, and an estimated position is represented by the coordinate $(x_e, y_e)$ in the two dimensions. Therefore, we can simply define the positioning error distance and represent by an Error Distance formula as follows:

$$error_{dis} = \sqrt{(x-x_e)^2 + (y-y_e)^2} \tag{33}$$

Meanwhile, in the three dimensions, the error distance is:

$$error_{dis} = \sqrt{(x-x_e)^2 + (y-y_e)^2 + (z-z_e)^2} \tag{34}$$

where $(x, y, z)$ is the real coordinate of the target node and $(x_e, y_e, z_e)$ is the estimated coordinate of the target node.

The experiment environment deployment is an indoor corridors which is $4.2m \times 6m$. There are six reference nodes and a coordinator. A laptop works as the center sever.

Through measuring the RSS within eighteen meters, the relationship between RSS and distance can be got by the curve fitting method and the result is shown in Figure 2, where the RSS has the good attenuation properties within 8.1m, so the logarithmic model also has been conducted in 8.1m. Thus we test three models, namely IEEE model, Log model and Log model (8.1m) in the following experiments.

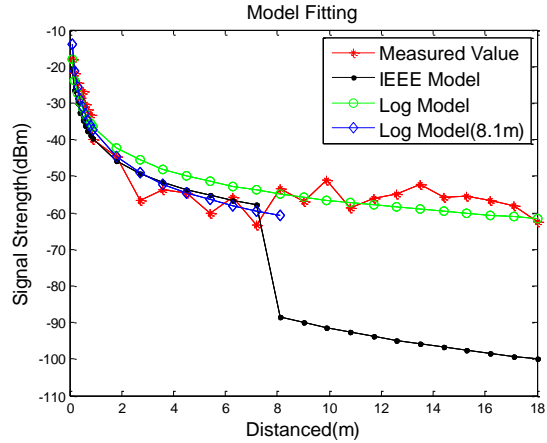

Figure 2: The curve fitting result.

## 5.1 Simulations

In the two dimension experiment, the nodes are deployed in the $5m \times 32m$ indoor corridor. The positions of 10 reference nodes are shown in Figure 3(a) and the coordinates are (0,0)、(0,5)、(8,0)、(8,5)、(16,0)、(16,5)、(24,0)、(24,5)、(32,0)、(32,5) respectively. The predefined movement is a straight line, from (1,3) to (31,3). Tracking experiment contains two parts. In the first part, there is only one mobile node moving along the trajectory. In the second, there are two mobile nodes, and another mobile node is moving from (1,2.5) to (31,2.5) at the same time. The two mobile nodes form an array.

The simulation results of Log model (8.1m) are shown in Figure 3(b). The tracking errors of Log model and IEEE model are summarized in Table.1. We can find that location accuracy has an obvious increase with the use of two nodes in two and there dimension case.

The 3D simulation experiment environment is deployed as shown in the Figure 3(c). There are eight reference nodes around the $8m \times 8m \times 8m$ indoor space. Both of the two mobile nodes' routes are curves. The mobile node 2 is the auxiliary node. In the process of moving, the relationship of the coordinates between two mobile nodes is $(\Delta x, \Delta y, \Delta z) = (0, 0, 0.5)$.

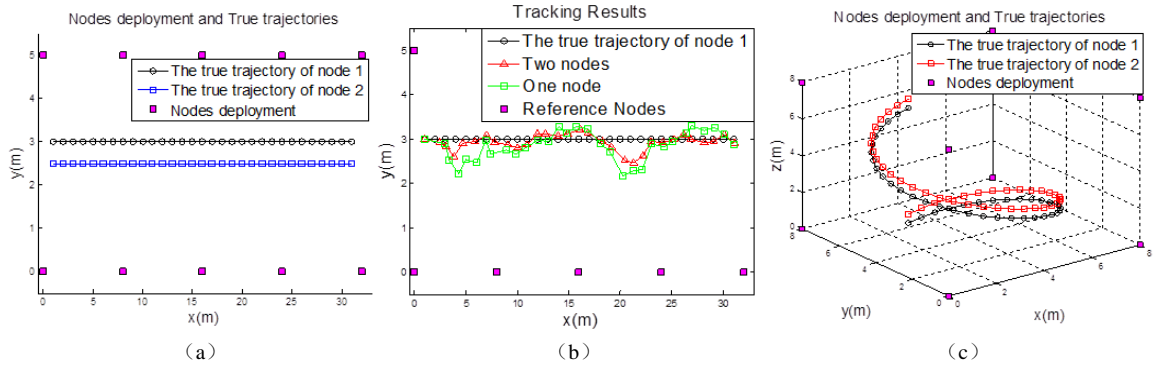

Figure 3:Simulations.

Table 1. Mean tracking Errors (m)

|  | MN number | Log model | Log model (8.1m) | IEEE model |
|---|---|---|---|---|
| 2D | One | 1.09785 | 0.37142 | 0.54343 |
|  | two | 0.55388 | 0.22817 | 0.39751 |
| 3D | One | 1.62755 | 1.11473 | 1.27925 |
|  | two | 1.25853 | 0.76551 | 1.05432 |

Through the above simulations, the following conclusions can be obtained. When the same model is used, the two nodes positioning errors is smaller than only one mobile node. When using the different models, the higher positioning accuracy can be achieved if the model is fit the actual environment. The Log model (8.1m) is the best model in the three models, and its positioning accuracy is the best. From the results in table 1, it can be seen that the proposed scheme has got a better precision improved.

## 5.2 Experiments

First, considering the two dimension, the nodes deployment and the predefined curve trajectory are shown in Figure 4(a). The reference nodes coordinates are (0,0), (0,3), (0,6), (4.2,0), (4.2,3), (4.2,6). The mobile node is moving from (1,1) to (3,5). The secondary node is 0.5 meters apart. And the coordinate relationship is $(\Delta x, \Delta y) = (0, 0.5)$.

The results of tracking errors with the three methods are shown in Figure 4(b)、(c) and (d). Their tracking errors are summarized in Table 2. We can find that location accuracy has an obvious increase with the use of two nodes in two dimension case.

Second, considering the three dimension, the nodes deployment and the predefined curve trajectory are shown in Figure 4(e). The reference nodes coordinates are (0,0,1), (0,3,1), (0,6,1), (4.2,0,1), (4.2,3,1), (4.2,6,1). The mobile node is moving from (1.8,0,1) to (3.6,0,1). The secondary node is 0.5 meters apart. And the coordinate relationship is $(\Delta x, \Delta y, \Delta z) = (0, 0, 0.5)$. The results of tracking errors with the three methods are shown in Figure 4(f)、(g) and (h). Their tracking errors are summarized in Table 2. We can find that location accuracy has an obvious increase with the use of two nodes in three dimension case.

The 2D and 3D results are summarized in Table 2. The two nodes positioning errors is smaller than only one mobile node. The results shows that the fitting model is closer to the actual environment, the positioning results are more accurate. For each kind of fitting method, the two mobile nodes positioning errors are smaller than one mobile node over 20cm.

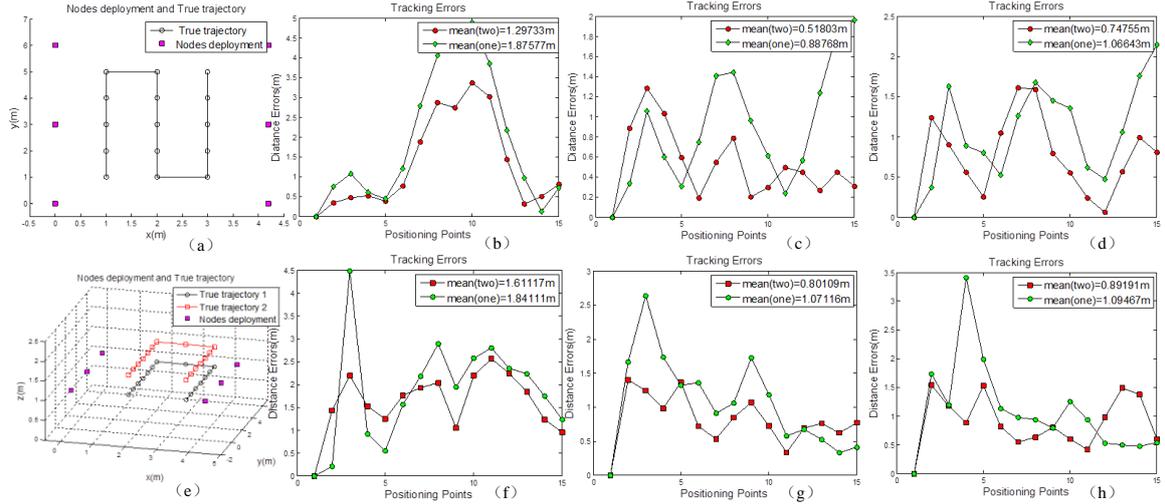

Figure 4: Experiments.

Table 2. Experiment Errors (m)

|  | MN number | Log model | Log model (8.1m) | IEEE model |
|---|---|---|---|---|
| 2D | One | 1.87577 | 0.88768 | 1.06643 |
|  | two | 1.29733 | 0.51803 | 0.74755 |
| 3D | One | 1.84111 | 1.07116 | 1.09467 |
|  | two | 1.61117 | 0.80109 | 0.89191 |

## 6 CONCLUSION

This paper presents a novel sensor fusion approach of Received Signal Strength (RSS) and IS, which deals with implementation of Zigbee Wireless Sensor Networks (WSN) using RSS among multiple mobile nodes (MN). It is different from traditional methods based on RSS. Indoor wireless localization suffers from severe multi-path fading and non-line-of-sight conditions. Through an efficient collaboration between two or more sensors, effectively exploits the RSS techniques. The unknown position is equipped with two or more mobile nodes. The range between two mobile nodes is fixed as priori. The attitude (roll, pitch, and yaw) of the mobile node are measured by inertial sensors (ISs). Then the angle and the range between any two nodes can be obtained. Theoretical analysis on localization distortion and Monte Carlo simulations shows that the proposed cooperative strategy of multiple Nodes with extended kalman filter achieves significantly higher localization accuracy compared to the existing systems, especially in heavily obstructed scenarios.

In this paper, we perform the tests in both the 2D and 3D indoor environment and apply the difference of experiment scenarios. The evaluation of the estimated error can analyze from the error distance between two points that is the estimated location and the true location by using Euclidean distance. From the observation of location estimation error, the 2D and 3D achieve the improvement of the positioning accuracy more than 20cm.

The corresponding simulations not only show the effectivity of the proposed method, but also show that the formation containing multiple mobile nodes will obtain better localization performance than single mobile node. Therefore, the proposed approach has great potential to help multiple mobile nodes moving in complex, limited indoor environment.

## 7  ACKNOWLEDGEMENTS

This work was financially supported by the National Natural Science Foundation of China under Grants 61371182, and Sichuan Science and Technology Support Project under Grants 2014GZ0037.
**REFERENCES**

[1] Xuxiang Fan, Gang Wang*, Jiachen Han, Yinghui Wang, A Background-Impulse Kalman Filter with Non-Gaussian Measurement Noises, IEEE Transactions on Systems, Man and Cybernetics: Systems, vol. 53, no. 4, pp. 2434-2443, April. 2023.

[2] Zhenyu Feng , Gang Wang , Bei Peng* , Jiacheng He , Kun Zhang , Distributed Minimum Error Entropy Kalman Filter, Information Fusion (2023), Volume 91, March 2023, Pages 556-565.

[3] Jiacheng He, Gang Wang*, Kui Cao, He Diao, Guotai Wang, Bei Peng, Generalized minimum error entropy robust learning, Pattern Recognition, 2023, vol. 135, pp. 109188.

[4] Jiacheng He, Hongwei Wang, Gang Wang, Shan Zhong, Bei Peng*, Minimum Error Entropy Rauch-Tung-Striebel Smoother, IEEE Transactions on Aerospace and Electronic Systems, doi: 10.1109/TAES.2023.3312057

[5] Shan Zhong, Bei Peng, Lingqiang Ouyang, Xinyue Yang, Hongyu Zhang, Gang Wang*，A Pseudolinear Maximum Correntropy Kalman Filter Framework for Bearings-only Target Tracking, IEEE Sensors Journal, vol. 23, no. 17, pp. 19524-19538, 1 Sept.1, 2023, doi: 10.1109/JSEN.2023.3283863

[6] Chen Liu, Gang Wang*, Xin Guan, Chutong Huang, Robust M-estimation-based Maximum Correntropy Kalman Filter, ISA transaction, 2023, vol. 136 , pp.198-209

[7] Xinyue Yang, Yifan Mu, Kui Cao, Mengzhuo Lv, Bei Peng,Ying Zhang, Gang Wang*, Robust kernel recursive adaptive filtering algorithms based on M-estimate, Signal Processing, 2023, vol. 207, pp. 108952.

[8] Gang Wang, Linqiang Ouyang, Linling Tong, Qiang Fang, Bei Peng*, A Recursive Least P-Order Algorithm Based on M-estimation in a non-Gaussian Environment, IEEE Transactions on Circuits and Systems II: Express Briefs, vol. 70, no. 8, pp. 2979-2983, Aug. 2023

[9] Xingli Zhou, Guoliang Li, Ziyi Wang, Gang Wang, Hongbin Zhang*，Robust hybrid affine projection filtering algorithm under α-stable environment, Signal Processing, 2023, vol. 208, pp. 108981.

[10]   Jiacheng He, Gang Wang, Huijun Yu, Junming Liu, Bei Peng*, Generalized Minimum Error Entropy Kalman Filter for Non-Gaussian Noise, ISA Transactions, 2023,  vol. 136, pp.663-675

[11]   Zhenyu Feng, Gang Wang, Bei Peng*, Jiacheng He, Kun Zhang, Novel robust minimum error entropy wasserstein distribution kalman filter under model uncertainty and non-gaussian noise, Signal Processing, Volume 203, 2023, pp. 108806.

[12]   Jiacheng He, Gang Wang, Xi Zhang, Hongwei Wang, Bei Peng*, Maximum total generalized correntropy adaptive filtering for parameter estimation, Signal Processing, Volume 203, 2023, pp. 108787
12